\documentclass[onecolumn,epjc3]{svjour3}
\usepackage{mathtext,bbm,amsmath,amsfonts,amssymb,indentfirst,syntonly,graphicx}
\smartqed  
\RequirePackage{graphicx}
\RequirePackage{subfigure}
\RequirePackage{xcolor}
\def\bc{\begin{center}}
\def\ec{\end{center}}
\def\be{\begin{eqnarray}}
\def\ee{\end{eqnarray}}

\journalname{Eur. Phys. J. C}

\begin{document}
\title{A unified description for dipoles of the fine-structure constant and SnIa Hubble diagram in Finslerian universe}
\author{Xin Li \thanksref{e1,addr1,addr3},
        Hai-Nan Lin \thanksref{e2,addr2},
        Sai Wang \thanksref{e3,addr3},
        Zhe Chang \thanksref{e4,addr2,addr3}
}
\thankstext{e1}{lixin1981@cqu.edu.cn}
\thankstext{e2}{linhn@ihep.ac.cn}
\thankstext{e3}{wangsai@itp.ac.cn}
\thankstext{e4}{changz@ihep.ac.cn}
\institute{Department of Physics, Chongqing University, Chongqing 401331, China\label{addr1}\and
           Institute of High Energy Physics, Chinese Academy of Sciences, Beijing 100049, China\label{addr2}\and
           State Key Laboratory Theoretical Physics, Institute of Theoretical Physics, Chinese Academy of Sciences, Beijing 100190, China\label{addr3}
}
\date{Received: date / Accepted: date}
\maketitle

\begin{abstract}
We propose a Finsler spacetime scenario of the anisotropic universe. The Finslerian universe requires both the fine-structure constant and accelerating cosmic expansion have dipole structure, and the directions of these two dipoles are the same. Our numerical results show that the dipole direction of SnIa Hubble diagram locates at $(l,b)=(314.6^\circ\pm20.3^\circ,-11.5^\circ\pm12.1^\circ)$ with magnitude $B=(-3.60\pm1.66)\times10^{-2}$. And the dipole direction of the fine-structure constant locates at $(l,b)=(333.2^\circ\pm8.8^\circ,-12.7^\circ\pm6.3^\circ)$ with magnitude $B=(0.97\pm0.21)\times10^{-5}$. The angular separation between the two dipole directions is about $18.2^\circ$.
\end{abstract}

\section{Introduction}

During the last decades, the standard cosmological model, i.e., the cold dark matter with a cosmological constant ($\Lambda$CDM) model \cite{Sahni,Padmanabhan} has been well established. It is consistent with several precise astronomical observations that involve Wilkinson Microwave Anisotropy Probe (WMAP) \cite{Komatsu}, Planck satellite \cite{Planck1}, Supernovae Cosmology Project \cite{Suzuki}, and so on. One of the most important and basic assumptions of the $\Lambda$CDM model is the cosmological principle, which states that the universe is homogeneous and isotropic on large scales. However, such a principle faces several challenges \cite{Perivolaropoulos}. The Union2 SnIa data hint that the universe has a preferred direction pointing to $(l,b)=(309^\circ,18^\circ)$ in the galactic coordinate system \cite{Antoniou}. Toward this direction, the universe has the maximum acceleration of expansion. Astronomical observations \cite{Watkins} found that the dipole moment of the peculiar velocity field on the direction $(l,b)=(287^\circ\pm9^\circ,8^\circ\pm6^\circ)$ in the scale of $50h^{-1}\rm Mpc$ has a magnitude $407\pm81~\rm km\cdot s^{-1}$. This peculiar velocity is much larger than the value $110~ \rm km\cdot s^{-1}$ constrained by WMAP5 \cite{WMAP5}. The recent released data of Planck Collaboration show deviations from isotropy with a level of significance ($\sim3\sigma$) \cite{Planck2}. Planck Collaboration confirms asymmetry of the power spectra between two preferred opposite hemispheres. These facts hint that the universe may have certain preferred directions.

Both the $\Lambda$CDM model and the standard model of particle physics require no variation of fundamental physical constants in principle, such as the electromagnetic fine-structure constant $\alpha_e=e^2/\hbar c$. Recently, the observations on quasar absorption spectra show that the fine-structure constant varies at cosmological scale \cite{Webb,King:2012}. Furthermore, in high redshift region ($z>1.6$), they have shown that the variation of $\alpha_e$ is well represented by an angular dipole model pointing in the direction $(l,b)=(330^\circ,-15^\circ)$ with level of significance ($\sim4.2\sigma$). Mariano and Perivolaropoulos \cite{Mariano} have shown that the dipole of $\alpha_e$ is anomalously aligned with corresponding dark energy dipole obtained through the Union2 sample. One possible reason of the variation of $\alpha_e$ is the variation of the speed of light, which means that Lorentz symmetry is violated on cosmological scale. The fact that the universe may have a preferred direction also means the isotropic symmetry of cosmology is violated. Also, dipole direction of the fine-structure constant are aligned with cosmological preferred direction. Such facts hint that both the two astronomical observations, cosmological preferred direction and variation of $\alpha_e$, may correspond to the same physical mechanism.

Finsler geometry is a possible candidate for investigating both the cosmological preferred direction and the dipole structure of the fine-structure constant. Finsler geometry \cite{Book by Bao} is a new geometry which involves Riemann geometry as its special case. Chern pointed out that Finsler geometry is just Riemann geometry without quadratic restriction, in his Notices of AMS. The symmetry of spacetime is described by the isometric group. The generators of isometric group are directly connected with the Killing vectors. It is well known that the isometric group is a Lie group in Riemannian manifold. This fact also holds in Finslerian manifold \cite{Deng}. Generally, Finsler spacetime admits less Killing vectors than Riemann spacetime does \cite{Finsler PF}. The causal structure of Finsler spacetime is determined by the vanishing of Finslerian length \cite{Pfeifer1}. And the speed of light is modified. It has been shown that the translation symmetry is preserved in flat Finsler spacetime \cite{Finsler PF}. Thus, the energy and momentum are well defined in Finsler spacetime.

{\bf The property of Lorentz symmetry breaking in flat Finsler spacetime makes Finsler geometry to be a possible mechanism of Lorentz violation \cite{Kostelecky:2011,Kostelecky:2012}. Historically, Bogoslovsky \cite{Bogoslovsky:1977a,Bogoslovsky:1977b,Bogoslovsky:1983,Bogoslovsky:2004,Bogoslovsky:2007} first suggested a Finslerian metric, i.e., $ds=(\eta_{\mu\nu}dx^\mu dx^\nu)^{(1-b)/2}(n_\rho dx^\rho)^b$, to investigate Lorentz violation. Here, $\eta_{\mu\nu}$ is Minkowski metric and $n_\rho$ is a constant null vector. Such metric involves Lorentz symmetry violation without violation of relativistic symmetry \cite{Bogoslovsky:2005,Bogoslovsky:2006}. The relativistic symmetry is realized by means of the 3-parameter group of generalized Lorentz boosts. Later on the results obtained in Bogoslovsky's works were mostly reproduced by Gibbons et al. \cite{Gibbons}, with the help of the techniques of continuous deformations of the Lie algebras and nonlinear realizations. Gibbons et al. have pointed out that the Bogoslovsky spacetime corresponds to General Very Special Relativity which generalizes Glashow's very special relativity \cite{Coleman:1997,Coleman:1999,Cohen:2006}. In the same work \cite{Gibbons} the 8-parametric isometry group \cite{Bogoslovsky:1977a,Balan:2012} of the Bogoslovsky spacetime was called $DISIM_b(2)$. Although the group $DISIM_b(2)$ is an 8-dimensional subgroup of the 11-dimensional Weyl group, pure dilations are not elements of $DISIM_b(2)$. The gravitational field equation in Finsler spacetime has been studied extensively \cite{Li:2010,Miron,Rutz1,Vacaru,Stavrinos:2009,Stavrinos:2010,Vacaru:1997,Vacaru:2001,Pfeifer,Basilakos:2013}. Models \cite{Chang:2013zwa,Chang:2013kss,Chang:2013vla,Chang:2013lxa,Chang:2014sjs,Chang:2014njh} based on Finsler spacetime have been developed to study the cosmological preferred directions.}

We suggested that the vacuum field equation in Finsler spacetime is equivalent to the vanishing of Ricci scalar \cite{Finsler Bullet}. The vanishing of Ricci scalar implies that the geodesic rays are parallel to each other. The geometric invariant of Ricci scalar implies that the vacuum field equation is insensitive to the connection, which is an essential physical requirement. The Schwarzschild metric can be deduced from a solution of our field equation if the spacetime preserves spherical symmetry. Supposing spacetime to preserve the symmetry of ``Finslerian sphere",  we found a non-Riemannian exact solution of the Finslerian vacuum field equation \cite{Finsler BH}. In this paper, following the similar approach of our previous work \cite{Finsler BH}, we present a modified Friedmann-Robertson-Walker (FRW) metric in Finsler spacetime, and then use it to study the cosmological preferred direction and the dipole structure of the fine-structure constant.

The rest of the paper is arranged as follows. In Section \ref{sec:anisotropy}, we briefly introduce the anisotropic universe in Finsler geometry. It can be seen easily that the speed of light in vacuum (so the fine-structure constant) is direction-dependent. In Section \ref{sec:field-equation}, we derive the gravitational field equation in Finslerian universe, and obtain the distance-redshift relation. In section \ref{sec:observation}, we use the Union2.1 compilation to derive the preferred direction of the universe. We find that it is obviously aligned with the dipole direction of the fine-structure constant. Finally, conclusions and remarks are given in Section \ref{sec:conclusion}.

\section{Anisotropic universe}\label{sec:anisotropy}

Finsler geometry is based on the so-called Finsler structure $F$ defined on the tangent bundle of a manifold $M$, with the property $F(x,\lambda y)=\lambda F(x,y)$ for all $\lambda>0$, where $x\in M$ represents position and $y\equiv dx/d\tau$ represents velocity. The Finslerian metric is given as \cite{Book by Bao}
\begin{equation}
 g_{\mu\nu}\equiv\frac{\partial}{\partial y^\mu}\frac{\partial}{\partial y^\nu}\left(\frac{1}{2}F^2\right).
\end{equation}
In physics, the Finsler structure $F$ is not positive-definite at every point of Finsler manifold. We focus on investigating Finsler spacetime with Lorentz signature. A positive, zero and negative $F$ correspond to time-like, null and space-like curves, respectively. For massless particles, the stipulation is $F=0$. The Finslerian metric reduces to Riemannian metric, if $F^2$ is quadratic in $y$. One non-Riemannian Finsler spacetime is Randers spacetime \cite{Randers}. It is given as
\begin{equation}\label{Randers form}
 F_{Ra}(x,y)\equiv \alpha(x,y)+\beta(x,y),
\end{equation}
where
\begin{eqnarray}
 \alpha (x,y)&\equiv&\sqrt{\tilde{a}_{\mu\nu}(x)y^\mu y^\nu},\\
 \beta(x,y)&\equiv& \tilde{b}_\mu(x)y^\mu,
\end{eqnarray}
and $\tilde{a}_{ij}$ is Riemannian metric. Throughout this paper, the indices are lowered and raised by $g_{\mu\nu}$ and its inverse matrix $g^{\mu\nu}$. And the objects that decorate with a tilde are lowered and raised by $\tilde{a}_{\mu\nu}$ and its inverse matrix $\tilde{a}^{\mu\nu}$.

In this paper, we propose an ansatz that the Finsler structure of the universe is of the form
\begin{equation}\label{FRW like}
F^2=y^ty^t-a^2(t)F_{Ra}^2,
\end{equation}
where we require the vector $\tilde{b}_i$ in $F_{Ra}$ is of the form $\tilde{b}_i=\{0,0,B\}$ and $B$ is a constant. Here, the Riemannian metric $\tilde{a}_{ij}$ of Randers space $F_{Ra}$ is set to be Euclidian, that is, $\tilde{a}_{ij}=\delta_{ij}$. Thus the Finslerian universe (\ref{FRW like}) returns to FRW spacetime while $B=0$. The above requirement for $\tilde{b}_i$ and $\tilde{a}_{ij}$ implies that the spatial part of the universe $F_{Ra}$ is a flat Finsler space, since all types of Finslerian curvatures vanish for $F_{Ra}$. The Killing equations of Randers space $F_{Ra}$ are given as \cite{Finsler PF,Finsler BH}
\begin{eqnarray}\label{Killing eq1}
L_V \alpha &=&\tilde{a}_{\mu\nu,\gamma}\tilde{V}^\gamma+\tilde{a}_{\gamma\nu}\tilde{V}^\gamma_{~,\mu}+\tilde{a}_{\gamma\mu}\tilde{V}^\gamma_{~,\nu}= 0,\\ \label{Killing eq2}
L_V \beta &=&\tilde{V}^\mu\frac{\partial \tilde{b}_\nu}{\partial x^\mu}+\tilde{b}_\mu\frac{\partial \tilde{V}^\mu}{\partial x^\nu}= 0,
\end{eqnarray}
where $L_V$ is the Lie derivative along the Killing vector $V$ and the comma denotes the derivative with respect to $x^\mu$. Noticing that the vector $\tilde{b}_i$ parallel to $z$-axis, we find from the Killing equations (\ref{Killing eq1},\ref{Killing eq2}) that there are four independent Killing vectors in Randers space $F_{Ra}$. Three of them represent the translation symmetry, and the rest one represents the rotational symmetry in $x-y$ plane. It means that the rotational symmetry in $x-z$ and $y-z$ plane are broken. This fact means that the Finslerian universe (\ref{FRW like}) is anisotropic.

To derive the relation between luminosity distance and redshift, first we need investigate the redshift of photon in Finslerian universe (\ref{FRW like}). The redshift of photon can be derived from the geodesic equation. The geodesic equation for Finsler manifold is given as\cite{Book by Bao}
\begin{equation}\label{geodesic}
\frac{d^2x^\mu}{d\tau^2}+2G^\mu=0,
\end{equation}
where
\begin{equation}\label{geodesic spray}
G^\mu=\frac{1}{4}g^{\mu\nu}\left(\frac{\partial^2 F^2}{\partial x^\lambda \partial y^\nu}y^\lambda-\frac{\partial F^2}{\partial x^\nu}\right)
\end{equation}
is called geodesic spray coefficients. It can be proved from the geodesic equation (\ref{geodesic}) that the Finsler structure $F$ is a constant along the geodesic. Plugging the Finsler structure (\ref{FRW like}) into the formula (\ref{geodesic spray}), we obtain that
\begin{eqnarray}\label{geodesic spray t}
G^0&=&\frac{1}{2}a\dot{a}F_{Ra}^2,\\
\label{geodesic spray i}
G^i&=&Hy^iy^0,
\end{eqnarray}
where the dot denotes the derivative with respect to time and $H\equiv\dot{a}/a$ is the Hubble parameter. Then, the geodesic equations in Finsler universe (\ref{FRW like}) are given as
\begin{eqnarray}\label{geodesic eq1}
\frac{d^2t}{d\tau^2}+a\dot{a}F_{Ra}^2&=&0,\\
\frac{d^2x^i}{d\tau^2}+2H\frac{dx^i}{d\tau}\frac{dx^0}{d\tau}&=&0.
\end{eqnarray}
In Finsler spacetime the null condition of photon is given as $F=0$. It is of the form
\begin{equation}\label{null condition}
\left(\frac{dt}{d\tau}\right)^2-a^2F^2_{Ra}=0.
\end{equation}
Plugging the null condition (\ref{null condition}) into the geodesic equation (\ref{geodesic eq1}), we obtain the solution
\begin{equation}
\frac{dt}{d\tau}\propto\frac{1}{a}.
\end{equation}
It yields that the formula of redshift $z$
\begin{equation}\label{redshift1}
1+z=\frac{c_0}{ca},
\end{equation}
where $c$ is the speed of light and the subscript zero denotes the quantities given at present epoch.

The recent Michelson-Morley experiment carried through by M\"{u}ller et al. \cite{Muller} gives a precise limit on Lorentz invariance violation. Their experiment shows that the change of resonance frequencies of the optical resonators is of this magnitude $|\delta\omega/\omega|\sim10^{-16}$. It means that the Minkowski spacetime describes well the inertial system on the earth. Thus, we must require no variation of the speed of light at the present epoch. In Finslerian universe, the local inertial system at large cosmological scale is built by the flat Finsler spacetime, namely,
\begin{equation}
F^2_f=y^ty^t-F_{Ra}^2.
\end{equation}
Thus, the radial speed of light at large cosmological scale can be derived from $F_f=0$. It is of the form
\begin{equation}\label{SOL}
c_r=\frac{1}{1+B\cos\theta},
\end{equation}
where $\theta$ denotes the angle with respect to $z$-axis. Then, plugging the equation (\ref{SOL}) into the formula (\ref{redshift1}), and noticing $c_{r0}=1$, we obtain
\begin{equation}\label{redshift2}
1+z=\frac{1+B\cos\theta}{a}.
\end{equation}

A direct deduction gives that the variation of the speed of light is the variation of the fine-structure constant. By making use of the formula (\ref{SOL}), we obtain the variation of the fine-structure constant
\begin{equation}\label{vari alpha}
\frac{\Delta \alpha_e}{\alpha_e}=-\frac{\Delta c_r}{c_{r0}}=B\cos\theta+O(b^2).
\end{equation}
Here, we suppose that the Finslerian parameter $B$ is a small quantity. The formula (\ref{vari alpha}) tells us that $\Delta \alpha_e/\alpha_e$ have a dipole distribution at cosmological scale, which is compatible with the observations on quasar absorption spectra \cite{Webb,King:2012}.

\section{Gravitational field equation in Finslerian universe}\label{sec:field-equation}

In Finsler geometry, there is a geometrical invariant quantity, i.e., Ricci scalar. It is of the form \cite{Book by Bao}
\begin{equation}\label{Ricci scalar}
Ric\equiv R^\mu_\mu=\frac{1}{F^2}\left(2\frac{\partial G^\mu}{\partial x^\mu}-y^\lambda\frac{\partial^2 G^\mu}{\partial x^\lambda\partial y^\mu}+2G^\lambda\frac{\partial^2 G^\mu}{\partial y^\lambda\partial y^\mu}-\frac{\partial G^\mu}{\partial y^\lambda}\frac{\partial G^\lambda}{\partial y^\mu}\right),
\end{equation}
where $R^\mu_{~\nu}=R^{~\mu}_{\lambda~\nu\rho}y^\lambda y^\rho/F^2$. Though $R^{~\mu}_{\lambda~\nu\rho}$ depends on connections, $R^\mu_{~\nu}$ does not \cite{Book by Bao}. The Ricci scalar only depends on the Finsler structure $F$ and is insensitive to connections. Plugging the geodesic coefficients (\ref{geodesic spray t},\ref{geodesic spray i}) into the formula of Ricci scalar (\ref{Ricci scalar}), we obtain that
\begin{eqnarray}\label{Ricci scalar1}
F^2Ric=-3\frac{\ddot{a}}{a}y^ty^t+(a\ddot{a}+2\dot{a}^2)F^2_{Ra}.
\end{eqnarray}
Here, we define the modified Einstein tensor in Finsler spacetime as
\begin{equation}\label{Einstein tensor}
G^\mu_\nu\equiv Ric^\mu_\nu-\frac{1}{2}\delta^\mu_\nu S,
\end{equation}
where the Ricci tensor is defined as \cite{Akbar}
\begin{equation}\label{Ricci tensor}
Ric_{\mu\nu}=\frac{\partial^2\left(\frac{1}{2}F^2 Ric\right)}{\partial y^\mu\partial y^\nu},
\end{equation}
and the scalar curvature in Finsler spacetime is given as $S=g^{\mu\nu}Ric_{\mu\nu}$. Plugging the equation of Ricci scalar (\ref{Ricci scalar1}) into the formula (\ref{Einstein tensor}), we obtain
\begin{eqnarray}\label{G t}
G^t_t&=&3H^2,\\
\label{G i}
G^i_j&=&\left(\frac{2\ddot{a}}{a}+H^2\right)\delta^i_j,
\end{eqnarray}

Following the similar approach in Ref. \cite{Finsler BH}, in order to construct a self-consistent gravitational field equation in Finsler spacetime (\ref{FRW like}), we investigate the covariant conserve properties of the modified Einstein tensor $G^\mu_\nu$. The covariant derivative of $G^\mu_\nu$ in Finsler spacetime is given as \cite{Book by Bao}
\begin{equation}\label{covariant der}
G^\mu_{\nu~|\mu}=\frac{\delta}{\delta x^\mu}G^\mu_\nu+\Gamma^\mu_{\mu\rho}G^\rho_\nu-\Gamma^\rho_{\mu\nu}G^\mu_\rho,
\end{equation}
where
\begin{equation}\label{delta der}
\frac{\delta}{\delta x^\mu}=\frac{\partial}{\partial x^\mu}-\frac{\partial G^\rho}{\partial y^\mu}\frac{\partial}{\partial y^\rho},
\end{equation}
and $\Gamma^\mu_{\mu\rho}$ is the Chern connection. Here, we have used `$|$' to denote the covariant derivative. The Chern connection can be expressed in terms of geodesic spray coefficients $G^\mu$ and Cartan connection $A_{\lambda\mu\nu}\equiv\frac{F}{4}\frac{\partial}{\partial y^\lambda}\frac{\partial}{\partial y^\mu}\frac{\partial}{\partial y^\nu}(F^2)$ \cite{Book by Bao}
\begin{equation}\label{chern connection}
\Gamma^\rho_{\mu\nu}=\frac{\partial^2 G^\rho}{\partial y^\mu \partial y^\nu}-A^\rho_{\mu\nu|\kappa}\frac{y^\kappa}{F}.
\end{equation}
Noticing that the modified Einstein tensor $G^\mu_\nu$ does not have $y$-dependence, and Cartan tensor $A^\rho_{\mu\nu}=A^i_{jk}$ (index $i,j,k$ run over $\theta,\varphi$), one can easily get that the Chern connection $\Gamma^\rho_{\mu\nu}$ equals to the Christoffel connection that deduced from FRW metric if $\Gamma^\rho_{\mu\nu}\neq\Gamma^i_{jk}$. By making use of this property and the formula of geodesic spray (\ref{geodesic spray t},\ref{geodesic spray i}), we find that
\begin{equation}
G^\mu_{\nu~|\mu}=0.
\end{equation}
Now, we have proved that the modified Einstein tensor is conserved in Finsler spacetime. Then, following the spirit of general relativity, we propose that the gravitational field equation in the given Finsler spacetime (\ref{FRW like}) should be of the form
\begin{equation}\label{field equation}
G^\mu_\nu=8\pi G T^\mu_\nu,
\end{equation}
where $T^\mu_\nu$ is the energy-momentum tensor. The volume of Finsler space \cite{Shen:2001} is generally different with the one of Riemann geometry. However, in terms of Busemann-Hausdorff volumm form, the volume of a close Randers-Finsler surface is the same with unit Riemannian sphere \cite{Shen:2001}. This is why we have used $\pi$ in the field equation (\ref{field equation}).

Since the modified Einstein tensor $G^\mu_\nu$ only depends on $x^\mu$, thus the gravitational field equation (\ref{field equation}) requires that the energy-momentum tensor $T^\mu_\nu$  depends on $x^\mu$ and contains diagonal components only. Therefore, we set $T^\mu_{\nu}$ to be the form as
\begin{equation}
T^\mu_\nu={\rm diag}(\rho,-p,-p,-p),
\end{equation}
where $\rho=\rho(x^\mu)$ and $p=p(x^\mu)$ are the energy density and the pressure density of universe, respectively.
Then, the gravitational field equation (\ref{field equation}) can be written as
\begin{eqnarray}\label{field eq t}
3H^2&=&8\pi G\rho,\\
\label{field eq i}
\frac{2\ddot{a}}{a}+H^2&=&-8\pi G p.
\end{eqnarray}
And the covariant conservation of energy-momentum tensor $T^\mu_{\nu~|\mu}$ gives
\begin{equation}\label{conservation T}
\dot{\rho}+3H(\rho+p)=0.
\end{equation}
Combining the equations (\ref{field eq t},\ref{conservation T}), we obtain
\begin{equation}\label{H2}
H^2=H_0^2(\Omega_{m0}a^{-3}+\Omega_\Lambda),
\end{equation}
where $H_0$ is Hubble constant, $\Omega_{\Lambda}\equiv8\pi G\rho_{\Lambda}/(3H_0^2)$ and $\Omega_{m0}\equiv8\pi G\rho_{m0}/(3H_0^2)$.

\section{Observational constraints on Finslerian universe}\label{sec:observation}

Here we focus on using the Union2.1 SnIa data \cite{Suzuki:2012} to study the preferred direction of the universe and constrain the magnitude of Finslerian parameter $B$. By making use of equations (\ref{null condition},\ref{redshift2},\ref{H2}), the luminosity distance in Finslerian universe is given as
\begin{equation}\label{lumin dis}
d_L=(1+z)r=\frac{1+z}{H_0}\int_0^z\frac{dz}{\sqrt{\Omega_{m0}(1+z)^3(1-3B\cos\theta)+1-\Omega_{m0}}},
\end{equation}
where $r=\sqrt{x^2+y^2+z^2}$ is the radial distance. To find the preferred direction in Finslerian universe, we perform a least-$\chi^2$ fit to Union2.1 SnIa data
\begin{equation}
\chi^2\equiv\sum\frac{(\mu_{\rm th}-\mu_{\rm obs})^2}{\sigma_\mu^2},
\end{equation}
where $\mu_{\rm th}$ is the theoretical distance modulus given by
\begin{equation}
\mu_{\rm th}=5\log_{10}\frac{d_L}{\rm Mpc}+25~.
\end{equation}
$\mu_{\rm obs}$ and $\sigma_\mu$, given by the Union2.1 SnIa data, denote the observed values of the distance modulus and the measurement errors, respectively. The least-$\chi^2$ fit of formula (\ref{lumin dis}) to the Union2.1 data shows that the preferred direction locates at $(l,b)=(314.6^\circ\pm20.3^\circ,-11.5^\circ\pm12.1^\circ)$, and the magnitude of anisotropy $B=(-3.60\pm1.66)\times10^{-2}$. The preferred direction is consistent with the dipole direction derived by reference \cite{Mariano}. Before using our model to fit the Union2.1 SnIa data , we have fixed the Hubble constant $H_0$ and $\Omega_{m0}$ to be $H_0=70.0~\rm km\cdot s^{-1}\cdot Mpc^{-1}$ and $\Omega_{m0}=0.278$, which are derived by fitting the Union2.1 data to the standard $\Lambda$CDM model.

\begin{figure}[htbp]
\centering
\includegraphics[scale=0.5]{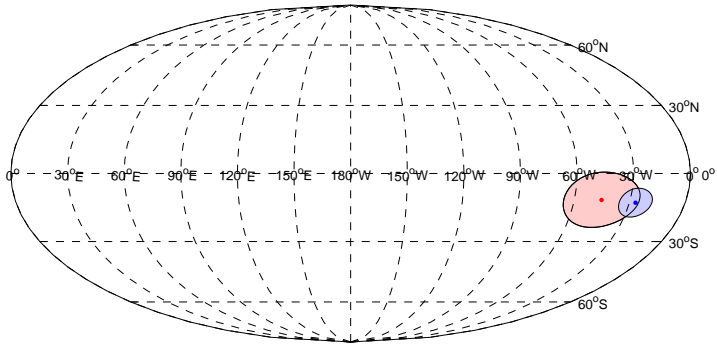}
\caption{The preferred directions in the galactic coordinate system. The red point locates at $(l,b)=(314.6^\circ\pm20.3^\circ,-11.5^\circ\pm12.1^\circ)$, which is obtained by fixing the parameters $\Omega_{m0}=0.278$ and $H_0=70.0~\rm km\cdot s^{-1}\cdot Mpc^{-1}$ and doing the least-$\chi^2$ fit to the Union2.1 data for formula (\ref{lumin dis}). The blue point locates at $(l,b)=(333.2^\circ\pm8.8^\circ,-12.7^\circ\pm6.3^\circ)$, which is obtained by the least-$\chi^2$ fit to the data of the fine-structure constant. The contours enclose 68\% confidence regions for the preferred directions. The angular separation between the two preferred directions is about $18.2^\circ$.}
\label{fig:Finslerian_dipole_f1}
\end{figure}

In Finslerian universe (\ref{FRW like}), the fine-structure constant has a dipole structure (see equation (\ref{vari alpha}) for details). We fit the data of quasar absorption spectra \cite{King:2012} obtained by Very Large Telescope (VLT) and Keck Observatory to formula (\ref{vari alpha}), and find the magnitude of dipole to be $B=(0.97\pm0.21)\times10^{-5}$, pointing towards $(l,b)=(333.2^\circ\pm8.8^\circ,-12.7^\circ\pm6.3^\circ)$ in the galactic coordinate system. We plot the preferred direction of the Union2.1 sample and the dipole direction of the fine-structure constant in the galactic coordinate system in Fig.\ref{fig:Finslerian_dipole_f1}.
We can see that they are consistent within $1\sigma$ uncertainty. The angular separation between the two directions is about $18.2^\circ$.

\section{Conclusions and Remarks}\label{sec:conclusion}

In this paper, we have suggested that the universe is Finslerian. The Finslerian universe (\ref{FRW like}) breaks the rotational symmetry in $x-z$ and $y-z$ plane, and modifies the speed of light at large cosmological scale. The preferred direction of Union2.1 SnIa sample and the dipole structure of the fine-structure constant are naturally given in Finslerian universe. Formula (\ref{vari alpha}) and (\ref{lumin dis}) show that the preferred directions of cosmic accelerating expansion and fine-structure constant both locate at the same direction. This fact is compatible with our numerical results. By applying our formula (\ref{lumin dis}) to Union2.1 SnIa data, we found a preferred direction $(l,b)=(314.6^\circ\pm20.3^\circ,-11.5^\circ\pm12.1^\circ)$. And by applying our formula (\ref{vari alpha}) to the data of quasar absorption spectra, we found a preferred direction $(l,b)=(333.2^\circ\pm8.8^\circ,-12.7^\circ\pm6.3^\circ)$. The angular separation between the two preferred directions is about $18.2^\circ$, which means that the two directions are compatible within $1\sigma$ uncertainty. However, our numerical results show that the anisotropic magnitude that correspond to Union2.1 SnIa data and the data of quasar absorption spectra are different. This fact contradicts with the prediction of Finslerian universe, which requires the absolute values of magnitude of the anisotropy should be the same. If the universe is Finslerian, such contradiction may be attributed to two reasons. One reason is the Finslerian parameter $B$ should be a function of redshift. Since the range of redshift of the two observational datesets are different. The other reason is that the Union2.1 SnIa data are not accurate enough. Our numerical results for the magnitude of Finslerian parameter $B$ show that the statistical significance of Union2.1 SnIa data is about $2\sigma$ confidence level, while the data of the fine-structure constant is about $4\sigma$ confidence level. Thus, further astronomical observations are needed in order to enhance the statistical significance in the future.

\begin{acknowledgements}
Project 11375203 and 11305181 supported by NSFC.
\end{acknowledgements}


\end{document}